# Topology-Enabled Self-Verification of Quantum Hall Resistance Standards

DaVonne Henry,[1#] Dominick S. Scaletta,[2#] Tamar M. Padwa,[3] Heather M. Hill,[1] Gretchen B. Ferguson,[4] Evelyn E. Hopkins,[4] Mattias Kruskopf,[5] Dean G. Jarrett,[1] David B. Newell,[1] and Albert F. Rigosi[1 *]

[1]*Physical Measurement Laboratory, National Institute of Standards and Technology (NIST), Gaithersburg, MD 20899, United States*

[2]*Department of Physics, Mount San Jacinto College, Menifee, California 92584, United States*

[3]*Department of Physics, Georgetown University, Washington, DC 20057, United States*

[4]*Department of Physics, Auburn University, Auburn, AL 36832, United States*

[5]*Physikalisch-Technische Bundesanstalt (PTB), Bundesallee 100, 38116 Braunschweig, Germany*

ABSTRACT: The scaling of multi-element quantized Hall array resistance standards necessitates rigorous validation that internal quantization has not broken down without relying on the verification of dissipationless edge state transport. One proposed solution is to use a Symmetric Precision Quantization Reliability criterion, a mathematical framework that statistically assures such validation strictly through outer terminal measurements of $C_n$-symmetric designs. This criterion essentially bounds the likelihood of hidden localized defects arising from quantum Hall breakdown and has potential uses in quantum computing and semiconductor testing but is presented in the context of its application toward quantum electrical standards.

[*] Email: afr1@nist.gov

[#] These authors contributed equally to this work.

## I. INTRODUCTION

The 2019 redefinition of the International System of Units (SI) marked a fundamental shift in precision measurement, anchoring base units directly to exact numerical values of fundamental physical constants. In quantum electrical metrology, the unit of resistance (ohm, Ω) is realized via the quantum Hall effect, which ties resistance to the von Klitzing constant $R_{\mathrm{K}} = \frac{h}{e^2}$, where $h$ is the Planck constant and $e$ is the elementary charge [1]. In recent history, some National Metrology Institutes' (NMI) primary resistance standards relied on the stable $i = 2$ Landau level in epitaxial graphene (EG) for realization ($R_{\mathrm{H}} = \frac{1}{2}\frac{h}{e^2} \approx 12906.4037\ \Omega$) [2] – [6].

Recognition of EG's technological maturity for standards manifested in 2026 with globally applicable technical guidelines establishing standardized measurement protocols for EG-based quantized Hall resistance (QHR) devices, released by the Bureau International des Poids et Mesures, an intergovernmental organization acting on global weights and measures [7]. Building on the success of devices with single Hall elements, recent attention has also focused on expanding the accessible range of quantized values through quantized Hall array resistance standards (QHARS) [8] – [12]. By interconnecting multiple Hall elements in series, parallel, or complex networks requiring mathematical star-mesh transformations [13] – [15], one can directly realize arbitrary and decadic quantized resistance values, eliminating the need for intermediate scaling steps and thus reducing the chain of traceability to $h$ and $e$ [16] – [17].

Before any international adoption of QHARS devices can occur, a major technical issue must first be resolved, namely, the verification that quantization has not locally broken down across all elements within an array. Failure or breakdown of quantization in a single element can corrupt the overall measured resistance without manifesting as an obvious deviation or anomaly during an arbitrary measurement, but attempting to test every element for quantization with the current guidelines of obtaining information about the longitudinal resistance of the element ($\rho_{xx}$) is unfeasible given the number of contacts necessary [7]. To overcome this challenge, rotationally symmetric topologies are proposed for EG-based QHARS devices to grant them self-diagnostic features. Coupled with the use of a Symmetric Precision Quantization Reliability (SPQR) criterion ($\Phi$), a statistical threshold of acceptance in lieu of the more exhaustive every-element test, these self-diagnostic device topologies offer a long-sought solution to the issue of QHARS verification. Beyond quantum electrical metrology, this criterion may also offer an alternative pathway for statistically assessing emerging quantum technologies, such as validation of topological qubit fidelities from input-output boundaries, or in semiconductor testing, where the topological fault-amplification principles used here may potentially be applied to atomic-scale node faults for high-confidence, low-pin-count automated testing.

## II. NUMERICAL AND EXPERIMENTAL METHODS

### A. Prediction and Simulation Techniques

QHARS device designs were mostly predicted manually with star-mesh transformations designed to target nominal decadal values of quantized resistance for future metrological applicability (values under $R_{\mathrm{H}}$ were designed on a case-by-case basis) [14]. Five quantized resistance outputs are explored in this work: 1 TΩ, 1 GΩ, 1 MΩ, 100 Ω, and about 8604 Ω (experimental device, $\frac{2}{3}R_{\mathrm{H}}$). All design outputs were simulated with LTspice to confirm the manually predicted values [18] – [19]. Figure 1 shows all five example values with designs utilizing varying degrees of rotational symmetry, as well as a tested experimental device.

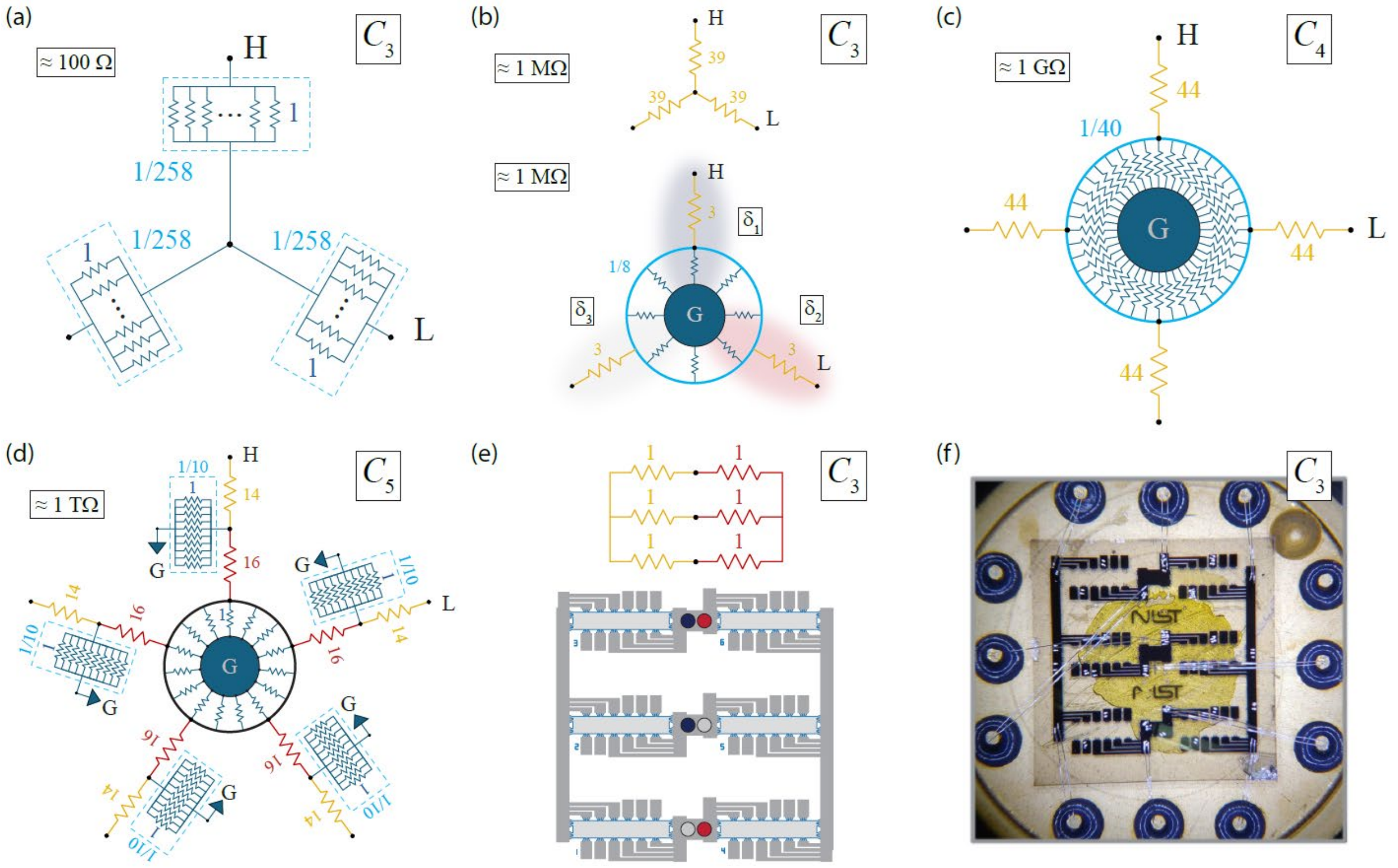

FIG. 1. (Color online) Schematics of 5 QHARS designs. (a) Approximately 100 Ω $C_3$-symmetric array featuring highly parallelized active arms. (b) Approximately 1 MΩ $C_3$-symmetric network, with two different approaches. The first approach uses series-only arrangements and the second utilizes star-mesh configurations. (c) Approximately 1 GΩ $C_4$-symmetric design that balances 44-element series arms with a 40-element parallel central grounded subarray (shunt). (d) Approximately 1 TΩ $C_5$-symmetric dual-stage star-mesh network. Each active arm contains a localized parallel shunt isolating two distinct series strings before connecting to the central secondary shunt going to ground. (e) Lithographic layout of a simpler $C_3$ ($\frac{2}{3}R_{\mathrm{H}}$ output, with three possible $R_{\mathrm{H}}$-valued cyclic measurements) design for validating SPQR criterion framework. The diagram labels the set of three cyclic two-terminal measurements. (f) Optical image of a physical $C_3$ device chip mounted and wire-bonded on a TO-8 carrier.

## B. Experimental Information

An experimental QHARS device was fabricated to demonstrate proof of principle of the SPQR criterion and uses monolayer EG grown on semi-insulating 4H-SiC substrates via polymer-assisted sublimation growth combined with face-to-graphite thermal annealing at 1900 °C in argon per previous work [17], [20]. The device is configured into six interconnected EG Hall elements joined by NbTiN superconducting leads [17], with charge carrier density tuned via covalent $Cr(CO)_3$ functionalization [21] – [22] (see Fig. 1 (e) and (f)). The QHARS device was characterized using Stanford Research Systems SR830 lock-in amplifiers in a cryogenic system at 1.5 K and an injected current of either 38.7 μA or 77.5 μA (see Acknowledgments).

Experimental observation windows are defined for three exemplary, separate measurement systems common in the field. The QHARS device was maintained at 1.5 K when tested with a Wheatstone-style bridge known as a direct current comparator (DCC) [23] and a separate cryogenic current comparator (CCC), the latter of which is generally more sophisticated a less user-friendly than a DCC. The CCC usually achieves uncertainties (noise floors) around the 1 nΩ/Ω level [24], whereas a DCC usually achieves slightly larger noise floors (near the 10 nΩ/Ω level). For quantized resistances several orders of magnitude higher than $R_H$, a dual source bridge is preferred for improved compatibility with the QHARS device (though not explicitly used for the above device), and it can achieve noise levels slightly under 100 nΩ/Ω [25] – [26].

To precisely determine the QHARS device values using the CCC bridge, 1 kΩ and 100 Ω standard resistors (Electro-Scientific Industries) were used. The standard resistors have a well-known linear drift rate and were calibrated about 100 days prior to the measurement of the QHARS device using a GaAs-AlGaAs quantized Hall resistance standard.

## III. DERIVATION OF THE SYMMETRIC PRECISION QUANTIZATION RELIABILITY CRITERION

To establish the Symmetric Precision Quantization Reliability (SPQR) criterion (Φ), which indicates a confidence level for acceptance that a QHARS has not experienced quantization breakdown, one may begin by defining the nature of an internal defect $\delta_i$, such that the resistance of the $i$-th quantum Hall element may be characterized as $R_i = R_H + \delta_i$. The following two known conditions about quantization breakdown will provide some guidance: (1) Thermodynamic irreversibility with regard to dissipation means that such a defect must only add resistance. Therefore, at the element level, $\delta_i \geq 0$. (2) The longitudinal resistivity exhibited by an element that has undergone a phase transition out of the quantum Hall effect is approximately defined as $\mu_i$; this has been reported as near 1 Ω [2], [27] – [29]. This value is used as the nominal value for the scale of the breakdown when computing the macroscopic analog $\mu_n$, to be described later. Since many elements can experience breakdown simultaneously, this framework uses the smallest possible value of $\mu_n$ since the objective is to claim with some level of confidence that the entire QHARS device (or some minimum number of subarrays of the device) does not exhibit localized quantization breakdown. That claim ultimately collapses when one element breaks down.

Since it may be infeasible to pinpoint the exact failed element, one must select a defect scalar that represents the smallest subsection of a device whose defective nature could be identified. At the same time, one may introduce the need for $n$-fold (rotational) symmetry to obtain a set of measurements that may be characterized as part of the cyclic group $C_n$. This cyclic nature will be essential for defining a probability that all measurements, if in strong agreement, are a result of artificial (not purely quantized) precision. Each subarray that undergoes rotation about a central node in its overall topology is defined as having resistance $R_n = R_{\text{nominal}} + \delta_n$ (where $\delta_n$ is a quantization error or defect resistance scalar). As in Fig. 1 (b), let $\vec{\delta}$ be the vector of these $n$ subarray defect resistances such that deviation in resistance of the $N$-th measurement configuration can be expressed as a combination of these defects (labeled the sensitivity matrix $\boldsymbol{S}$): $\Delta R_N = \sum_{j=1}^{N} S_{N,j}\delta_j$. If $N$ distinct measurements are performed in an $n$-fold symmetric topology (where $N \geq n$), then the SPQR criterion should intuitively approach 1 with larger $N$, indicating better QHARS quantization confidence. More information on the transformation of the element-scale breakdown parameter $\mu_i$ and defect scalar $\delta_i$ to the macroscopic subarray breakdown parameter $\mu_n$ and defect $\delta_n$ may be found in the Supplemental Material [30].

When assigning a probability distribution to a variable with a known expected characteristic scale, the principle of maximum entropy dictates the use of the least biased distribution [31]. Since $\delta_n$ can include elements both in series and in parallel, $\delta_n$ may take on positive or negative values, thereby making the most mathematically rigorous distribution the Laplace distribution [31] (more information on the appropriateness of this distribution is in the Supplemental Material [30]). Assuming the $n$ subarrays of an $n$-fold rotationally symmetric array fail independently, the joint probability density function $f(\vec{\delta})$ for the entire macroscopic defect vector $\vec{\delta}$ is the product of their individual Laplace distributions:

$$f(\vec{\delta}) = \prod_{j=1}^{n} \frac{1}{2\mu_n} \exp\left(-\frac{|\delta_j|}{\mu_n}\right) \tag{1}$$

Equation 1 then simplifies to:

$$f(\vec{\delta}) = \frac{1}{(2\mu_n)^n} \exp\left( -\frac{1}{\mu_n} \sum_{j=1}^{n} |\, \delta_j | \right)$$

(2)

The probability of artificiality $P(A)$ is thus the probability that an internal defect state exists and that it lands inside some observation window $\epsilon$, corresponding to the standard deviation of the *N* measurements. To find *P(A),* Eq. 2 must be integrated over the abstract region of $\vec{\delta}$-space that satisfies the observation, namely $\Omega_\delta$. To provisionally accept a subarray as sufficiently quantized, the deviation of the measurement must satisfy: $|\Delta R_N| \leq \epsilon$ for all *N* measurement configurations (excludes sources of deviation that may be accounted for in an uncertainty budget such as errors present in the measuring system). Geometrically, this restricts the acceptable observation space to an $n$-dimensional hypercube, which can be expressed as: $\Delta \vec{R} \in [-\epsilon, \epsilon]^n$. Furthermore, for an internal defect to be classified as an artificiality, the macroscopic defect vector $\vec{\delta}$ must map into this observation. The permissible integration region $\Omega_\delta$ is therefore defined entirely by the structural mapping constraint: $\boldsymbol{S}\delta \in [-\epsilon, \epsilon]^n$

The integral for *P(A)* may now be expressed as:

$$P(A) = \int_{\Omega_\delta} f\,(\vec{\delta})\, d^n\delta = \int_{\Omega_\delta} \frac{1}{(2\mu_n)^n} \exp\left( -\frac{1}{\mu_n} \sum_{j=1}^{n} |\, \delta_j | \right) d^n\delta$$

(3)

More information on the determination of $\epsilon$, $\mu_n$, and the defect vector $\vec{\delta}$ can be found in the Supplemental Material [30]. The integral in Eq. 3 may be evaluated straightforwardly for particular real-world scenarios, which may allow for a rectangular matrix $\boldsymbol{S}$ (if the number of measurements *N* exceeds the order of rotational symmetry *n*). For changing variables, one may transform the coordinates from the hidden defect space ($\vec{\delta}$) into the measurable terminal space ($\Delta\vec{R}$). The linear transformation is $\Delta\vec{R} = \boldsymbol{S}\vec{\delta}$, which means $\vec{\delta} = (\boldsymbol{S}^T\boldsymbol{S})^{-1}\boldsymbol{S}^T\Delta\vec{R}$. The Jacobian of this transformation is $(\boldsymbol{S}^T\boldsymbol{S})^{-1}\boldsymbol{S}^T$ and the differential volume element becomes:

$$d^n\delta = \frac{1}{\sqrt{\det(\boldsymbol{S}^T\boldsymbol{S})}}\, d^n(\Delta R)$$

(4)

And by substitution, the exact form of the probability of artificiality becomes the following, where the matrix 1-norm appears in the exponential term:

$$P(A) = \frac{1}{\sqrt{\det(\boldsymbol{S}^T\boldsymbol{S})}(2\mu_n)^n} \int_{\mathcal{P}} \exp\left(-\frac{1}{\mu_n} \|(\boldsymbol{S}^T\boldsymbol{S})^{-1}\boldsymbol{S}^T\Delta\vec{R}\|_1\right) d^n(\Delta R)$$

(5)

The integration bounds are represented as $\mathcal{P}$, a convex polytope defined by intersection of the experimental measurement constraints. Supposing that the standard deviation of the *N* experimental resistance measurements is several orders of magnitude smaller than the characteristic scale for a quantum Hall subarray that has broken down, i.e. at least one element that has undergone a phase transition ($\epsilon \ll \mu_n$), the maximum possible value of any term inside the polytope $\mathcal{P}$ is restricted by $\epsilon$, making the exponent in Eq. 5 effectively near-zero. The integral over the bounds now becomes the geometric volume $V_{\mathcal{P}}$:

$$P(A) \approx \frac{1}{\sqrt{\det(\boldsymbol{S}^T\boldsymbol{S})}(2\mu_n)^n} \times V_{\mathcal{P}}$$

(6)

The volume is less than or equal to the volume of the full hypercube (in the event there are no additional constraints): $V_{\mathcal{P}} \leq V_{\epsilon} = (2\epsilon)^n$. Therefore, the conservative upper bound becomes:

$$P(A) \leq \frac{1}{\sqrt{\det(\boldsymbol{S}^T\boldsymbol{S})}}\left(\frac{\epsilon}{\mu_n}\right)^n \tag{7}$$

By taking the maximum volume of the hypercube, one establishes a worst-case scenario, allowing the SPQR criterion to be mathematically conservative:

$$\Phi \approx 1 - \frac{1}{\sqrt{\det(\boldsymbol{S}^T\boldsymbol{S})}}\left(\frac{\epsilon}{\mu_n}\right)^n \tag{8}$$

Depending on the criterion strength one is using, one may affirm or deny the earlier provisional acceptance of the QHARS as sufficiently quantized; that is, if a specific SPQR confidence level is not reached, one may reverse the supposition that the QHARS is sufficiently quantized to a high degree of probability. It should be noted that the SPQR criterion does not claim to rule out systematic offsets, so a rigorous uncertainty budget for sources of error should be compiled prior to approaching this analysis. Example calculations, though summarized in the next section, may be found in more detail in the Supplemental Material [30].

## IV. APPLICATION OF THE SPQR CRITERION

### A. 5σ Confidence and Reliability Efficiency

The next step is to apply the SPQR criterion framework to several examples, including those that represent recent and conventional methods used in quantum electrical metrology. The four example topologies in Fig. 1 (a) through (d) are used, and specific evaluations of each case are available in the Supplemental Material [30]. Evaluating Eq. 8, which varies from design to design (regardless of intended nominal value), yields the calculated results in Fig. 2, showing how each of the four cases' SPQR criteria behaves as a function of order of rotational symmetry (and for the 1 MΩ case, the star-mesh design with fewer elements is used).

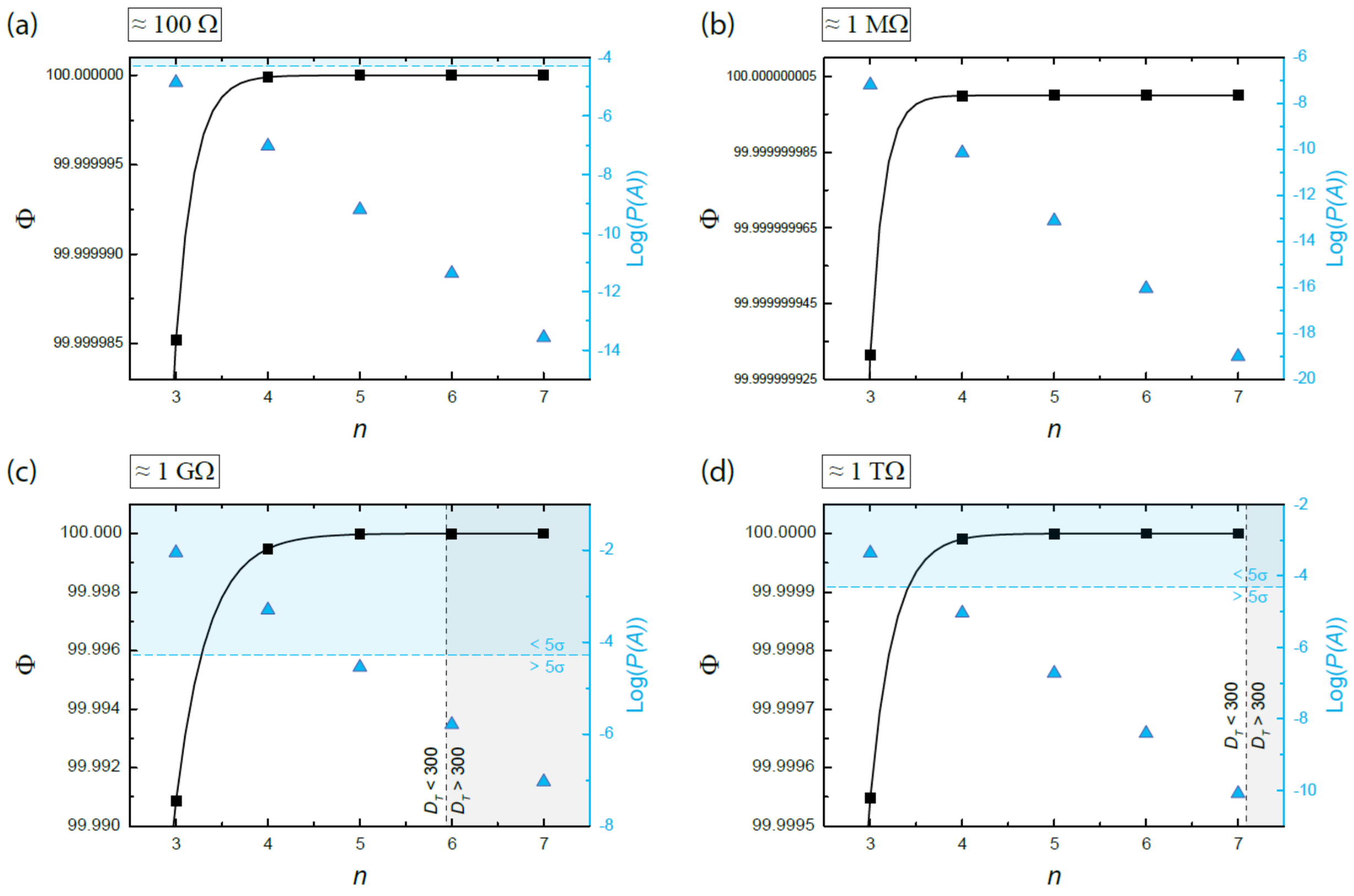

FIG. 2. (Color online) Evaluation of Eq. 8 for each of the following cases' SPQR criteria (left vertical axis and any marker in black or gray) and logarithm of the probability of artificiality (right vertical axis and any marker in cyan), both plotted as a function of order of rotational symmetry: (a) 100 Ω, (b) 1 MΩ (star-mesh configuration, not series-only configuration), (c) 1 GΩ, and (d) 1 TΩ. The dashed black line (and shaded gray areas) show the point beyond which higher SPQR criteria require an element count near or beyond current fabrication capabilities ($D_T$).

For the left and right vertical axes of Fig. 2, an imposed and arbitrary limitation are considered. The first, as a black dashed line for the left vertical axis, is an engineering limit of several hundred elements (where element count is labeled as $D_T$ per prior work) that has been recently demonstrated [32], though it may be possible to increase this limit with improvements in fabrication processes. That limit provides guidance on how much rotational symmetry one could leverage for self-diagnosing QHARS devices. The second limit, a cyan dashed line for the right vertical axis, is the level of confidence where one is satisfied with the probability that the full array has not experienced quantization breakdown, above which (cyan region) one is no longer in the proposed 5σ confidence regime. Though arbitrary, a 5σ SPQR criterion is proposed as a quantitative consistency test designed to indicate a likelihood that a device has not undergone quantization breakdown, mainly because the use of this high level of confidence is also seen in particle physics for major discoveries, such as the Higgs boson [33].

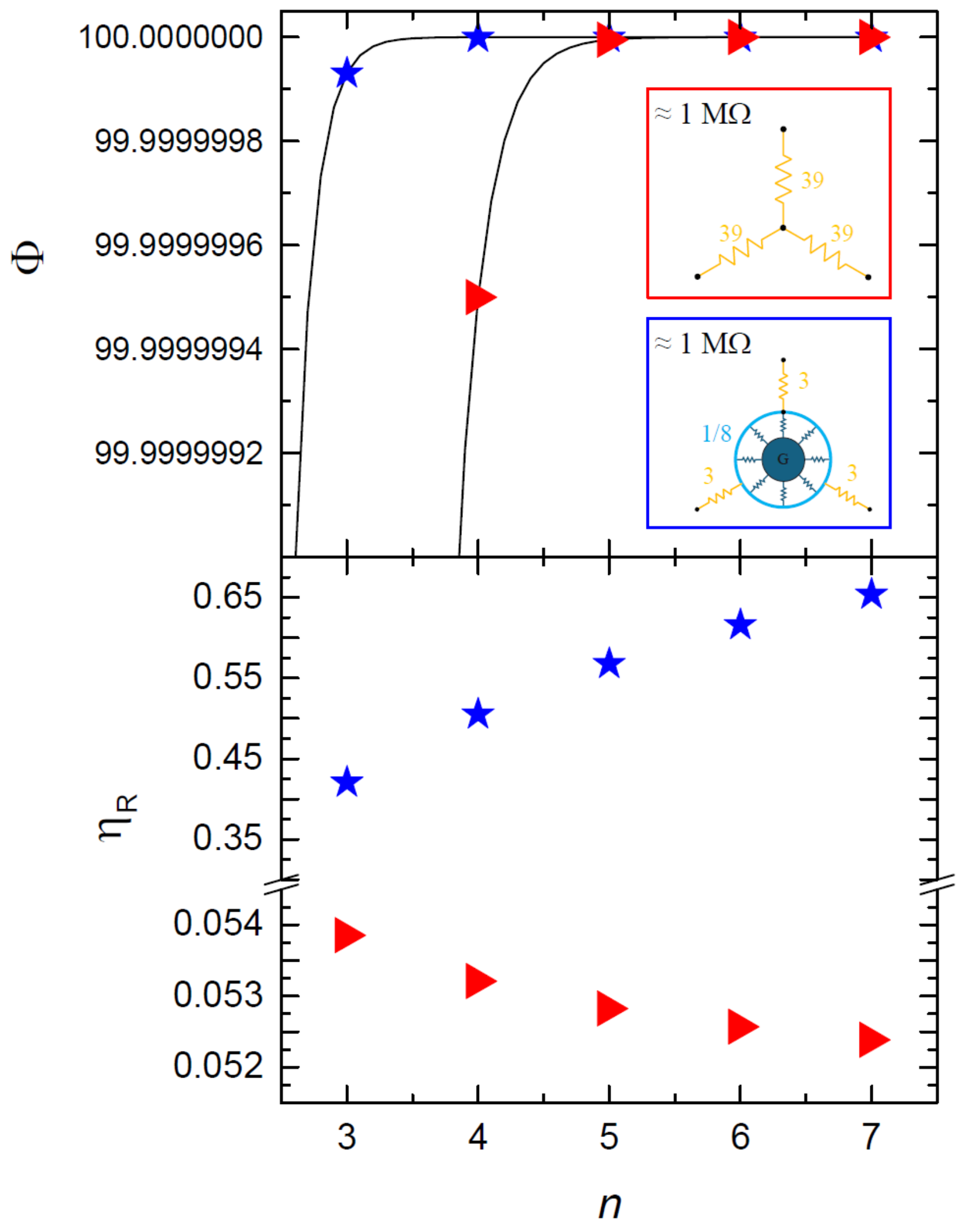

FIG. 3. (Color online) (top) Comparison of the SPQR criterion behavior between series-only topology (red triangles) and a corresponding star-mesh topology (blue stars) for the same exact value of quantized resistance (1 MΩ). (bottom) The reliability efficiency ($\eta_R$), shown as a function of order of rotational symmetry for two distinct 1 MΩ topologies, indicates that star-mesh topologies should be considered far more heavily than series-only topologies due to a much greater utilization of device footprint.

These arbitrary constraints provide some guidance as to what levels of rotational symmetry may be most beneficial for certain magnitudes of electrical resistance. A clearer example of these benefits is provided in Fig. 3, which compares the SPQR criterion behavior between series-only topology and a corresponding star-mesh topology for the same exact value of quantized resistance. What emerges is a more rapidly converging confidence for the star-mesh topology, as seen by the blue star calculations in Fig. 3 (top panel). Further, from an efficiency standpoint, it appears that the gains in SPQR criterion strength may not be worth the additional elements one must fabricate for the series-only topologies.

To quantify how well a specific topology utilizes its physical footprint in terms of potential gain in SPQR criterion strength, one can define the reliability efficiency ($\eta_R$) and requires both the probability of artificiality and element count for the full QHARS device ($D_T$, as used in Ref. [14]):

$$\eta_R = \frac{-\log_{10}(P(A))}{D_T} \tag{9}$$

Equation 9 thus allows one to more carefully analyze Fig. 1 (b), which showed two ways of achieving 1 MΩ (either with conventional series-only arms or star-mesh arms). Assuming the same 10 $\text{n}\Omega/\Omega$ bridge noise floor, the series-only design requires 117 elements and the star-mesh design requires only 17 elements. In the bottom panel of Fig. 3, one now observes that there is sufficient mathematical justification for utilizing star mesh transformations to achieve higher resistances since it reduces the number of elements drastically, increasing one's reliability efficiency.

### B. Experimental Demonstration

To reconcile the use of the SPQR criterion and how metrology measurements have been made in the past, one may turn to two major demonstrations. The first will be with the use of the QHARS from Fig. 1 (e-f). The second demonstration uses example single Hall element measurements and is shown in detail in the Supplemental Material. Figure 4 (a) shows the three-dimensional topological drawing of the QHARS, showing $n$ = 3 and $N$ = 3 and maps the measurements onto the real device layout. Lock-in amplifier measurements in Fig. 4 (b) as a function of magnetic flux density show the quantization of the three cyclic measurements as well as the singular full-array value of $\frac{2}{3}R_{\mathrm{H}}$, necessitating a measuring system of higher precision.

Precision measurements were performed with a DCC and a CCC and are shown in Fig. 4 (c) and (d), all performed at 9 T. The first step was to find the SPQR criterion of this array, whose detailed calculation is in the Supplemental Material. Based on the standard deviation of the 3 cyclical DCC measurements (shaded cyan region) in Fig. 4 (c) (all color coded to the pairs of two-terminal measurements in Fig. 4 (a)), one finds the observation window $\epsilon_{QHARS\,Device}$ and thus the SPQR criterion $\Phi \approx 1 - \frac{1}{2}\left(\frac{6.45\times10^{-4}\Omega}{0.25\ \Omega}\right)^3 \approx 0.999999991\ldots$, which is significantly stronger than 5σ (≈ 99.99994 %). The single circular data point in Fig. 4 (c) is an example CCC measurement of the middle pair to confirm that both measurement systems are agreeing to a reasonable quantity. A second current is used for two pairs just to highlight the benefits gained in precision when higher currents are used.

With this confidence metric now signifying an array the is unlikely to have broken down quantization, one need not rely on the knowledge of the longitudinal resistances of every element when measuring the full value of the QHARS device ($\frac{2}{3}R_{\mathrm{H}}$). But to further strengthen the propositions herein, a comprehensive assessment of each element's longitudinal resistivity is provided in the Supplemental Material [30]. Figure 4 (d) shows six CCC measurement sets for $\frac{2}{3}R_{\mathrm{H}}$, two for each of three calibrated standard 100 Ω resistors. The determined error bars account for the type A ($k$ = 1) uncertainties of the measurements. In short, once the QHARS is verified, one can measure other output values, if they are available, with the same confidence.

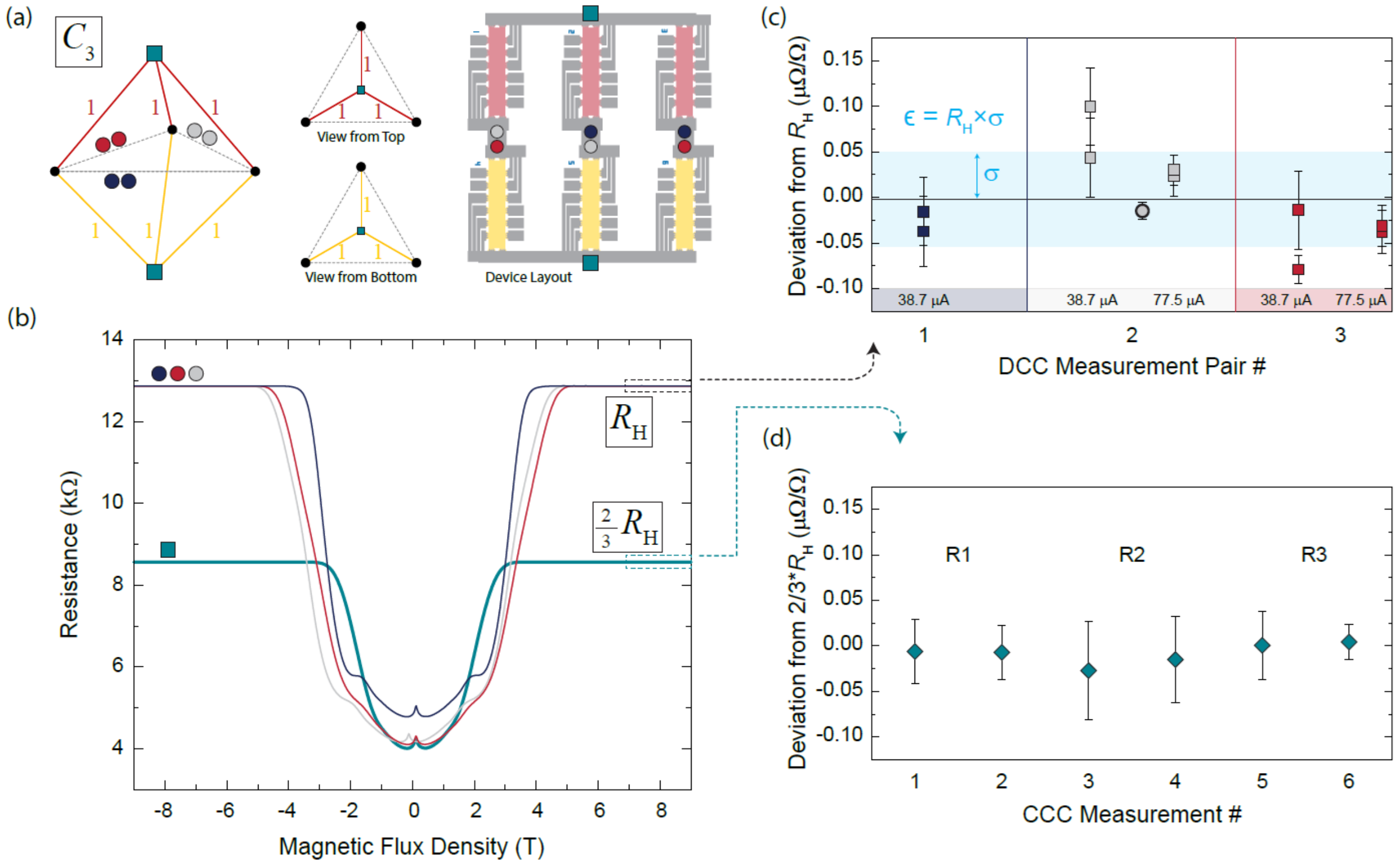

FIG. 4. (Color online) (a) Schematic and drawing of the three cyclic measurements shown for a QHARS (circular gray, blue, and red pairs of two-terminal measurements) and a singular full-array value of $\frac{2}{3}R_{\mathrm{H}}$ (square turquoise pair). (b) Magnetic flux density sweep for the corresponding measurements drawn in (a). (c) DCC data are shown for each of the three colors of possible cyclic measurements, with the shaded cyan region indicating standard deviation of the measurements, ultimately needed for evaluating the SPQR criterion for this QHARS device. (d) Once the device has been validated, one can measure other values, if available and not cyclic. A CCC was used to measure the full array value of $\frac{2}{3}R_{\mathrm{H}}$ and enjoys the benefit of the same confidence of quantization established by the DCC measurements performed on different array configurations. The CCC was used to perform two sets of measurements for each of the three calibrated standard 100 Ω resistors.

Ultimately, this ability to measure a QHARS device in a cyclical fashion helped validate the concept of the SPQR criterion. Furthermore, upon seeing the scales of the SPQR criterion for metrological measurements (both here and in the SM [30]), the proposal of 5σ appears reasonable, especially since it is also commensurate with confidence in other physical discoveries. Beyond metrology, this criterion may provide a broadly applicable framework for the statistical assessment and validation of emerging quantum technologies and other advanced systems, including high-confidence testing at increasingly complex and microscopic scales.

## V. CONCLUSIONS

This work demonstrates that by designing devices with $C_n$-symmetric topologies, one can leverage the SPQR criterion to statistically validate a lack of macroscopic array quantization breakdown strictly through two-terminal measurements. Experimental validation with a QHARS device confirmed that this framework reasonably achieves a $5\sigma$ confidence level, which is the proposed level of confidence for the metrology community. Constructive future steps could combine cyclic measurements with a few select targeted internal measurements, such as characterizing one Hall element per arm. A sampling of an individual longitudinal, Hall, and contact resistance may help constrain common-mode contributions that elude detection and thus affect the entire array.

The SPQR criterion may also be applicable to assessments of emerging quantum technologies, such as validation of qubit functionality from external boundaries or in semiconductor testing by means of topological fault tracking. This criterion need not be the only proposed method for utilizing mathematical principles for drastically simplifying engineering requirements, but at present, it remains the only formally proposed method.

## ACKNOWLEDGMENTS

Unless otherwise noted, NIST work was funded solely by the United States Government. The authors declare no competing interest. For strongly fruitful discussions on future applicability, the authors wish to thank member institutions of the Consultative Committee for Electricity and Magnetism (CCEM) and its Task Group on Graphene for Resistance Standards (all part of the International Bureau of Weights and Measures (BIPM, France)), with specific gratitude to H. Scherer from Physikalisch-Technische Bundesanstalt (PTB, Germany), M. Taupin and W. Poirier from Laboratoire National de Métrologie et d'Essais (LNE, France), L. Callegaro from Istituto Nazionale di Ricerca Metrologica (INRIM, Italy), H. He from Research Institutes of Sweden (RISE), S. Giblin, S. Rozhko, and A. Tzalenchuk from the National Physical Laboratory (NPL, United Kingdom), T. Oe from National Metrology Institute of Japan (NMIJ), D.-H. Chae from Korea Research Institute of Standards and Science (KRISS), and Prof. dr. ir. Gert Rietveld from Van Swinden Laboratorium (VSL, Netherlands).

The authors also wish to express their appreciation to P. Barbara at Georgetown University and D. Haddad, F. Siefert, Y. Yang, A. Panna, N. T. M. Tran, A. M. Possolo, and C. Clark at NIST for fruitful discussions. D. Henry would like to thank the National Research Council's Research Associateship Program for the opportunity.

Commercial equipment, instruments, and materials are identified in this paper in order to specify the experimental procedure adequately. Such identification is not intended to imply recommendation or endorsement by the National Institute of Standards and Technology or the United States Government, nor is it intended to imply that the materials or equipment identified are necessarily the best available for the purpose.

# Supplemental Material

# Topology-Enabled Self-Verification of Quantum Hall Resistance Standards ($C_n$-Symmetric Topologies)

DaVonne Henry,[1#] Dominick S. Scaletta,[2#] Tamar M. Padwa,[3] Heather M. Hill,[1] Gretchen B. Ferguson,[4] Evelyn E. Hopkins,[4] Mattias Kruskopf,[5] Dean G. Jarrett,[1] David B. Newell,[1] and Albert F. Rigosi[1 †]

*[1]Physical Measurement Laboratory, National Institute of Standards and Technology (NIST), Gaithersburg, MD 20899, United States*

*[2]Department of Physics, Mount San Jacinto College, Menifee, California 92584, United States*

*[3]Department of Physics, Georgetown University, Washington, DC 20057, United States*

*[4]Department of Physics, Auburn University, Auburn, AL 36832, United States*

*[5]Physikalisch-Technische Bundesanstalt (PTB), Bundesallee 100, 38116 Braunschweig, Germany*

Table of Contents:

---

[#] These authors contributed equally to this work.

**Table of Definitions and High-Level Description of the SPQR Criterion**

The Symmetric Precision Quantization Reliability (SPQR) criterion is a percentage confidence level that a given quantized Hall array resistance standard (QHARS) topology exists such that none of its elements have experienced a quantum phase transition away from the quantized Hall effect. The definitions of significant variables include:

[1] $\Phi$ - Symmetric Precision Quantization Reliability (SPQR) criterion, acting as a statistical threshold of acceptance or confidence level that all QHARS device elements have not exited the quantum Hall state.

[2] $n$ - Order of rotational symmetry for a device's topology (cyclic group $C_n$).

[3] $\delta_n$ - The macroscopic quantization error, or defect resistance scalar, for a single subarray. The full vector is $\vec{\delta}$.

[4] $\mu_n$ - Macroscopic subarray breakdown parameter.

[5] $\epsilon$ - The observation window, corresponding to the standard deviation of the $N$ measurements.

[6] $N$ - Number of distinct measurements performed in an $n$-fold symmetric topology.

[7] $\Delta R_N$ - The deviation in resistance of the $N$-th measurement configuration.

[8] $S$ - Sensitivity matrix containing elements $S_{N,j}$, which expresses the deviation in measurement as a combination of defects.

[9] $P(A)$ - The probability of artificiality, defined as the probability that an internal defect state or lack of quantization exists and lands inside the observation window $\epsilon$.

[10] $\eta_R$ - Reliability efficiency, which quantifies how efficiently a specific topology utilizes its physical footprint in terms of the potential gain in SPQR criterion strength.

[11] $D_T$ - The total element count for the full QHARS device.

## 1. Additional Experimental and Theoretical Details

Measurements were performed for an array in the main text using both a direct current comparator (DCC) and cryogenic current comparator (CCC). A dual source bridge only applies for higher resistance and so was not utilized in practice for this specific array (valued at $\frac{2}{3}R_{\mathrm{H}}$). Graphene areas were verified by optical and electrical methods described in prior work [1, 3, 4].

For the DCC, resistance measurements were conducted using a Measurements International (MI) 6020Q Automated Resistance Bridge. The device under test was an $8.60\,\mathrm{k\Omega}$ nominal resistor (QHARS device from the main text) and was measured against a $1000.000330\,\Omega$ calibrated reference standard (Electro-Scientific Industries). The bridge was operated in a 1:10 range configuration with a $I_s = 0.5\,\mathrm{mA}$ applied to the reference standard. Hardware settings included an analog-to-digital setting of 10 and a measurement settling delay of 12 s. A total of 125 data points were collected, with the final 100 points used for statistical analysis.

For the CCC, measurements were performed using a configuration with a primary winding $N_1$ of 1376 turns and a secondary winding $N_2$ of 16 turns. The primary current ($I_1$) was driven at an amplitude of $58.11\,\mu\mathrm{A}$ on a $75\,\mu\mathrm{A}$ range, with a negative feed-in polarity and the SQUID feed-in set to off at $-52.0\,\mathrm{dB}$. The secondary current ($I_2$) was set on a $20\,\mathrm{mA}$ range with a $5.0\,\mathrm{mA}$ amplitude, maintaining the superconducting quantum interference device (SQUID) feed-in active at $-44.5\,\mathrm{dB}$ and a target voltage delta $\Delta(I_2 \cdot R_2)$ of $1.0000\,\mathrm{V}$. Data acquisition was executed with an integration time of 6 time-base multiples and 100 samples per half-cycle, discarding the first 50 samples while retaining all trailing samples. The detection electronics operated in AMP mode featuring an amplifier gain of 300,000 (negative gain with a $1.65\,\mathrm{k\Omega}$ amplifier resistance), a flux-locked loop feedback resistance of 9.10 kΩ, and positive SQUID gain.

On the issue of selecting the Laplace distribution, which also matches the ISO/IEC Guide 98-3:2008/Suppl 1:2008, we wish to explain that the probability density function $f(\delta)$ in this framework is not used to model the empirical variance of our measurement data (which is captured separately by the observation window $\epsilon$). It instead models the unobservable, theoretical state space of hidden internal quantization defects. We are thus constrained to physical priors, namely, that we know the expected characteristic scale of a failed element ($\mu_i \approx 1\ \Omega$), but we do not possess a statistically well-behaved variance ($\sigma^2$) for these rare macroscopic faults. In the context of quantum electrical metrology, adopting a Gaussian distribution would aggressively suppress the mathematical likelihood of severe, singular macroscopic faults due to its $e^{-x^2}$ tail decay, whereas the Laplace distribution accommodates heavier tails ($e^{-|x|}$), providing a conservative upper bound for $P(A)$.

To further explain the joint probability density function, in statistical mechanics, a state space must be defined containing every possible configuration of a system, and then a probability must be assigned to each state. Therefore, $f(\delta)$ maps this defect space of the entire QHARS device. It answers the following question: "What is the exact probability that the whole design exists in a specific configuration wherein macroscopic errors exist?" Before looking at the whole device, one must first focus on just one symmetric arm (subarray). This arm is assigned a Laplace distribution, whose shape indicates that when one has many measurements in strong agreement, the most likely physical scenario (an Occam's razor approach) is that the arm is fully quantized ($\delta_j = 0$). If a defect does exist, it could manifest as a positive or negative macroscopic shift, with the probability of that shift occurring decaying exponentially as the shift gets larger.

The width of the distribution is determined by the macroscopic breakdown parameter $\mu_n$. If $\mu_n$ is massive (like in the 1 TΩ network), the distribution is very wide, meaning large macroscopic errors are highly probable if a microscopic fault exists. Mathematically, the probability of a specific error $\delta_j$ in a single arm is $\frac{1}{2\mu_n}\exp\left(-\frac{|\delta_j|}{\mu_n}\right)$.

The next step is to scale up to the entire device. The array has $n$ independent, rotationally symmetric arms. Recall from elementary physics that with two non-interacting particles, the probability of finding particle A in state $X$ and particle B in state $Y$ is simply the product of their individual probabilities. This exact same logic may be used to assume that a quantum breakdown in arm 1 is completely physically independent of a breakdown in arm 2, and because the failures are independent, to find the joint probability of the entire device's state, one multiplies the individual probabilities of all $n$ arms together, leading to Eqs. 1 and 2 of the main text:

$$f(\delta) = \prod_{j=1}^{n} \frac{1}{2\mu_n} \exp\left(-\frac{|\delta_j|}{\mu_n}\right) = \frac{1}{(2\mu_n)^n} \exp\left(-\frac{1}{\mu_n}\sum_{j=1}^{n} |\delta_j|\right) \tag{S1}$$

Equation S1 looks similar to a Boltzmann distribution ($e^{-E/kT}$), and so one can think of the sum term, $\sum|\delta_j|$, as the total defect energy of the entire QHARS device. The equation thus states that the probability of the entire QHARS device existing in one specific defective state drops exponentially based on the sum total of all the macroscopic errors present across the structure. Overall, this approach mathematically models the entire $n$-dimensional volume of every possible way the device can fail. The SPQR criterion then proceeds to calculate what fraction of that failure volume can hide inside the measurement noise limit ($\epsilon$) (since anything outside that noise limit can presumably be measured and thus potentially accounted for), and the larger that fraction, the lower the confidence in one's QHARS quantization.

## 2. Discussion on $\vec{\delta}$

This section will focus on demonstrating how one translates the defect vector of a single element to the corresponding macroscopic version. Since that translation is dependent on the exact layout that is selected, this example will focus on the 1 GΩ design presented in the main text (Fig. 1 (c), where each of the four branches may be labeled as North, East, South, and West). Recall that $\delta_i$ captures the small, localized deviations or non-idealities within individual components of the measurement network.

To begin, the high-dimensional internal defect space must be projected onto a reduced *n*-dimensional manifold. First define a strictly additive macroscopic segment defect, $\delta_n$, for each branch. From the terminal to a central node (and/or central ground), the effective resistance of a branch (such as the North quadrant) is modeled as $R_{\text{North}} = R_{\text{North, nominal}} + \delta_{\text{North}}$. Here, $\delta_{\text{North}}$ is an encapsulation of every microscopic degree of freedom within that branch. Recall that $R_{\text{elem}} = R_H + \delta_i$ - the macroscopic segment is then governed by a linear approximation of this 1 GΩ network's internal transfer function below (which expands $\Delta R_N = \sum_{j=1}^{N} S_{N,j}\delta_j$, where $N$ refers to the total number of measurements, with our focus on the measurement example in the main text between North and East):

$$\Delta R_{\text{eq, North-East}} = \left(\frac{\partial R_{\text{eq}}}{\partial R_{\text{series}}}\right)\Delta R_{\text{series}} + \left(\frac{\partial R_{\text{eq}}}{\partial R_{\text{parallel}}}\right)\Delta R_{\text{parallel}} \tag{S2}$$

And recalling the star-mesh transformation for $R_{\text{eq}}$ yields a conventional Wye-Delta formula:

$$R_{\text{eq}} = R_{\text{series}} + R_{\text{series}} + \frac{(R_{\text{series}})^2}{R_{\text{parallel}}} \tag{S3}$$

Substituting the first-order summations for the series arm and the parallel shunt, the complete micro-to-macro mapping expands into this form:

$$\Delta R_{\text{eq, North-East}} = \left(2 + \frac{2R_{\text{series}}}{R_{\text{parallel}}}\right)\left(\sum_{j=1}^{40} \delta_{i,\text{series},j}\right) - \left(\frac{R_{\text{series}}^2}{R_{\text{parallel}}^2}\right)\left(\frac{1}{44^2}\sum_{k=1}^{44} \delta_{\mu,\text{parallel},k}\right) \tag{S4}$$

This procedure maps the complex physical failure modes of the device into a single aggregated degree of freedom per branch. Note that the analysis in this section is not required for the general use of the SPQR criterion; it is provided for purposes of edification.

## 3. Determination of $\epsilon$

The observation window $\epsilon$ is expressed in ohms (Ω) and represents the standard deviation of a set of $N$ measurements made. In the case of ideal measurements, $\epsilon$ should reflect the noise floor of the instrumentation. So, for instance, at 1 MΩ, assuming an ideal measurement and a relative standard deviation minimum of 10 nΩ/Ω (i.e., by use of a CCC), one obtains:

$$\epsilon_{1\,M\Omega} = (10^6\ \Omega) \times 10^{-8}\Omega/\Omega = 0.01\ \Omega$$

And for the experimental setup at 1 GΩ, the scaling is intuitive:

$$\epsilon_{1\,G\Omega} = (10^9\ \Omega) \times 10^{-7}\Omega/\Omega = 100\ \Omega$$

For 100 Ω (assumes a 1 nΩ/Ω noise floor with a CCC) and 1 TΩ (assumes a 100 nΩ/Ω noise floor with a DSB), we have the following exemplary values of $\epsilon$:

$$\epsilon_{100\,\Omega} = (100\ \Omega) \times 10^{-9}\Omega/\Omega = 10^{-7}\Omega$$

$$\epsilon_{1\,T\Omega} = (10^{12}\ \Omega) \times 10^{-7}\Omega/\Omega = 10^{5}\Omega$$

Within the framework of the SPQR criterion, $\epsilon$ abstractly defines the acceptance boundary in $N$-dimensional measurement space (where $n$ is the order of rotational symmetry for a specific QHARS topology). Two things should be noted here: First, suppose $\epsilon$ (the spread of the measurements) is large. It then follows that evaluating the SPQR criterion yields a higher probability of artificiality, meaning that one's confidence that a QHARS device exhibits quantization intuitively decreases. Second, $\Delta R_N$ depends on $R_{\text{nominal}}$, which may or may not include errors that create offsets. Ultimately, the quantization probability mostly gives weight to the deviation of the measurements ($\epsilon$) as opposed to offsets that may be common to all data (and thus accountable in an uncertainty budget).

## 4. Discussion on the Breakdown Parameter $\mu_n$

The loss of quantization is a relatively abrupt phase transition, and when the critical current is exceeded, mechanisms like inter-Landau-level scattering cause the system to rapidly heat and break down, making the longitudinal resistance jump suddenly by several orders of magnitude beyond the ohm range. This allows the value of 1 ohm to be a conservative representation of the loss of quantization ($\mu_i$), and therefore, we define $\mu_n$ to be the macroscopic signal that quantization has fundamentally collapsed somewhere along the active measurement path (presumably, and as a stringent bound, due to a single element failure).

For context, recall that the defect vector $\vec{\delta}$ is an $n$-dimensional vector representing the entire rotationally symmetric network: $\vec{\delta} = [\delta_1, \delta_2, \ldots, \delta_n]^T$ Each individual scalar component (e.g., $\delta_1$) represents the macroscopic defect localized to a single arm. However, when we define the Laplace probability distribution for that specific arm's defect, $f(\delta_1) \propto \exp(-|\delta_1|/\mu_n)$, the macroscopic breakdown parameter $\mu_n$ reflects the entire two-terminal measurement, including the second arm and the shared central ground (in the 1 GΩ case of Fig. 1 (c) in the main text), and can be calculated using the chain rule to translate a microscopic elemental fault ($\mu_i \approx 1\ \Omega$) into a measurable shift in resistance. Let $R_i$ be the specific quantum Hall element experiencing this fault. Then $\mu_n$ is the total derivative of the measured equivalent resistance ($R_{\text{eq}}$) with respect to that specific faulty element:

$$\mu_n = \mu_i \left| \frac{dR_{\text{eq}}}{dR_i} \right|$$

$$\mu_n = \mu_i \left| \frac{\partial R_{\text{eq}}}{\partial R_{\text{arm}}} \times \frac{\partial R_{\text{arm}}}{\partial R_{\text{seg}}} \times \frac{\partial R_{\text{seg}}}{\partial R_i} \right| \tag{S5}$$

To find $\mu_n$ systematically for any topology and layout, one expands the total derivative into three distinct topological layers within which the fault could exist: (1) highest level, (2) arm level, and (3) element level. More terms for additional layers of topological complexity may be added, but since we explore values at the edge of application (1 TΩ), such additional layers and terms beyond what is in Eq. S5 are unlikely to be necessary.

The first term describes how the active measurement loop responds to a change in one of its arms (almost always two arms in this work), the second term describes how a single (presumably faulty) arm's total resistance changes if one of its internal macroscopic segments (e.g., a specific series string or a parallel bank) experiences a breakdown. If the topology is simple (e.g., an arm of just a single string of elements in series), this derivative is simply 1. The third term translates the single faulty quantum Hall element into a shift of its immediate segment (if there are multiple segments in an arm). The use of this approach will be seen in more detail in the full QHARS calculations in a later section.

One thing to note here is that $\mu_n$ is intended to represent a physically meaningful quantity as opposed to a historical metrological compliance limit. For instance, one may call an element not fully quantized if it exhibits a longitudinal resistance in the mΩ range, but such a designation is rather arbitrary compared to a phase transition.

## 5. Example QHARS Calculation in Full

Starting with the probability of artificiality $P(A)$:

$$P(A) = \frac{1}{\sqrt{\det(\boldsymbol{S}^T\boldsymbol{S})}(2\mu_n)^n} \int_{\mathcal{P}} \exp\left(-\frac{1}{\mu_n} \|\boldsymbol{S}^{-1}\Delta\vec{R}\|_1\right) d^n(\Delta R) = \ldots..$$

$$\ldots = \frac{1}{\sqrt{\det(\boldsymbol{S}^T\boldsymbol{S})}(2\mu_n)^n} \int_{\mathcal{P}} \exp\left(-\frac{1}{\mu_n}\sum_{j=1}^{n}\left|\sum_{k=1}^{N}(\boldsymbol{S}^{+})_{jk}\Delta R_k\right|\right) d^n(\Delta R)$$

(S6)

Additional context: When the total number of distinct measurements $N$ exceeds the $n$-fold rotational symmetry of the array ($N > n$) (and assuming that $\epsilon \ll \mu_n$; that is, at least one order of magnitude), the system is overdetermined and the $N \times n$ sensitivity matrix $\boldsymbol{S}$ becomes rectangular. Consequently, the standard inverse is replaced by the pseudoinverse $\boldsymbol{S}^{+} = (\boldsymbol{S}^T\boldsymbol{S})^{-1}\boldsymbol{S}^T$, and the Jacobian scaling factor of the linear transformation is governed by the square root of the Gram determinant, $\sqrt{\det(\boldsymbol{S}^T\boldsymbol{S})}$. Because the $N$ terminal measurements are physically constrained by only $n$ internal degrees of freedom, the permissible experimental polytope $\mathcal{P}$ represents an $n$-dimensional manifold embedded within the $N$-dimensional measurement space.

<u>For 100 Ω</u> (main text Fig. 1 (a)):

The $C_3$ symmetric design has three arms where each must have a resistance of about 50 Ω. With 258 Hall elements in parallel (temporarily labeled $L$ elements in parallel) within each of the three arms, one obtains:

$$R_{\text{arm}} = \frac{R_H}{258} \approx 50.025\ \Omega$$

This yields a total device element count of 774 elements, arguably within standard lithographic limits. To get the macroscopic breakdown parameter $\mu_n$, recall Eq. S5, and using the fact that $R_{\text{eq}} = R_{\text{arm-A}} + R_{\text{arm-B}}$, $R_{\text{arm-A}} = R_{\text{seg}}$ (since there are no further complications in the topology of each arm), and $R_{\text{seg}} = \left[\frac{L}{R_i}\right]^{-1}$:

$$\mu_n = \mu_i \left| \frac{\partial R_{\text{eq}}}{\partial R_{\text{arm-A}}} \times \frac{\partial R_{\text{arm-A}}}{\partial R_{\text{seg}}} \times \frac{\partial R_{\text{seg}}}{\partial R_i} \right|$$

$$\mu_n = \mu_i \left[ 1 * 1 * (-1)\left(\frac{L}{R_i}\right)^{-2} * (-1)\left(\frac{1}{R_i}\right)^{2} \right]$$

$$\mu_n \approx \frac{\mu_i}{L^2} = \frac{1.00\ \Omega}{258^2} \approx 15\ \mu\Omega$$

One may proceed with the three-measurement suite (arms 1-2, 2-3, 3-1), obtaining the sensitivity matrix ($\boldsymbol{S}$) and its Jacobian scale factor:

$$\boldsymbol{S} = \begin{bmatrix} 1 & 1 & 0 \\ 0 & 1 & 1 \\ 1 & 0 & 1 \end{bmatrix}, \quad |\det(\boldsymbol{S})| = 2$$

Now use the result for the SPQR criterion (Eq. 8 in the main text):

$$\Phi \approx 1 - \frac{1}{\sqrt{\det(\boldsymbol{S}^T\boldsymbol{S})}} \left(\frac{\epsilon}{\mu_n}\right)^n$$

$$\Phi \approx 1 - \frac{1}{2}\left(\frac{0.1\ \mu\Omega}{15\ \mu\Omega}\right)^3 \approx 0.99999985\ ...$$

And this result is slightly stronger than 5σ (≈ 99.99994 %).

<u>For 1 MΩ</u> (main text Fig. 1 (b)):

The $C_3$ symmetric design utilizes a simpler Wye-Delta configuration with each arm having three Hall elements in series and all three arms sharing a grounded shunt with eight elements in parallel. A measurement between any two outer terminals while grounding the central shunt yields approximately 1 MΩ.

To get $\mu_n$, recall Eq. S5 and use $R_{\text{eq}} = R_{\text{A}} + R_{\text{B}} + \frac{R_{\text{A}} R_{\text{B}}}{R_{\text{shunt}}}$, $R_{\text{A}} = R_{\text{seg}}$ (since there are no further complications in the topology of each arm), and $R_{\text{seg}} = R_i + 2R_H$ (since, under ideal conditions $R_{\text{seg}} = 3R_H$, but one of the elements $R_i$ is failing). Assuming the fault is in Arm A in the measurement loop (Arm B will yield an identical analysis):

$$\mu_n = \mu_i \left| \frac{\partial R_{\text{eq}}}{\partial R_{\text{A}}} \times \frac{\partial R_{\text{A}}}{\partial R_{\text{seg}}} \times \frac{\partial R_{\text{seg}}}{\partial R_i} \right|$$

$$\mu_n = \mu_i \left[ \left(1 + \frac{3R_H}{R_H/8}\right) * 1 * 1 \right] \approx 25\ \Omega$$

Before continuing, a new feature of implementing star-mesh configurations requires us to analyze both the arm and shunt separately since we do not know where the single faulty element is. Suppose it's in the shunt – then the partial derivative changes:

$$\mu_n = \mu_i \left| \frac{\partial R_{\text{eq}}}{\partial R_{\text{shunt}}} \times \frac{\partial R_{\text{shunt}}}{\partial R_{\text{seg}}} \times \frac{\partial R_{\text{seg}}}{\partial R_i} \right|$$

$$\mu_n = \mu_i \left[ \left(-\frac{(3R_H)(3R_H)}{(R_H/8)^2}\right) * 1 * \left(\frac{1}{8}\right)^2 \right] \approx 9\ \Omega$$

We select the smallest value of $\mu_n$ to set a lower bound for the SPQR criterion. The three-measurement suite is the same as the previous example: $|\det(\boldsymbol{S})| = 2$, and the SPQR criterion is:

$$\Phi \approx 1 - \frac{1}{2}\left(\frac{0.01\ \Omega}{9\ \Omega}\right)^3 \approx 0.999999999314 \ldots$$

And this result is stronger than 5σ (≈ 99.99994 %).

For 1 GΩ (main text Fig. 1 (c)):

The $C_4$ symmetric design also utilizes a simpler Wye-Delta configuration with each arm having 44 Hall elements in series and all three arms sharing a grounded shunt with 40 elements in parallel. A measurement between any two outer terminals while grounding the central shunt yields approximately 1 GΩ.

To get $\mu_n$, recall Eq. S5 and use $R_{\text{eq}} = R_{\text{A}} + R_{\text{B}} + \frac{R_{\text{A}}R_{\text{B}}}{R_{\text{shunt}}}$, $R_{\text{A}} = R_{\text{seg}}$ (since there are no further complications in the topology of each arm), and $R_{\text{seg}} = R_i + 43R_H$ (since, under ideal conditions $R_{\text{seg}} = 44R_H$, but one of the elements $R_i$ is failing). Assuming the fault is in Arm A in the measurement circuit (Arm B will yield an identical analysis):

$$\mu_n = \mu_i \left| \frac{\partial R_{\text{eq}}}{\partial R_{\text{A}}} \times \frac{\partial R_{\text{A}}}{\partial R_{\text{seg}}} \times \frac{\partial R_{\text{seg}}}{\partial R_i} \right|$$

$$\mu_n = \mu_i \left[ \left( 1 + \frac{44R_H}{R_H/40} \right) * 1 * 1 \right] \approx 1761\ \Omega$$

Again, if the shunt is the source of a fault, then the partial derivative changes:

$$\mu_n = \mu_i \left| \frac{\partial R_{\text{eq}}}{\partial R_{\text{shunt}}} \times \frac{\partial R_{\text{shunt}}}{\partial R_{\text{seg}}} \times \frac{\partial R_{\text{seg}}}{\partial R_i} \right|$$

$$\mu_n = \mu_i \left[ \left( -\frac{(44R_H)(44R_H)}{(R_H/40)^2} \right) * 1 * \left( \frac{1}{40} \right)^2 \right] \approx 1936\ \Omega$$

We select the smallest value of $\mu_n$ to set a lower bound for the SPQR criterion.

The four-measurement suite can be set up in the following manner (we chose *N* to be 4 to match *n*, the order of rotational symmetry, for keeping a square matrix, but having a larger *N* is now possible and would only improve the SPQR criterion). Also note that for even numbered orders, the final measurement should not complete the cyclic pattern for symmetry reasons:

$$\begin{bmatrix} \Delta R_1 \\ \Delta R_2 \\ \Delta R_3 \\ \Delta R_4 \end{bmatrix} = \begin{bmatrix} 1 & 1 & 0 & 0 \\ 0 & 1 & 1 & 0 \\ 0 & 0 & 1 & 1 \\ 1 & 0 & 1 & 0 \end{bmatrix} \begin{bmatrix} \delta_N \\ \delta_E \\ \delta_S \\ \delta_W \end{bmatrix}$$

The determinant evaluates precisely to $\det(\boldsymbol{S}) = -2$. Because $|\det(\boldsymbol{S})| \neq 0$, the matrix is full-rank and strictly invertible, guaranteeing that no combination of branch defects can linearly camouflage themselves across the suite of measurements. The SPQR criterion is:

$$\Phi \approx 1 - \frac{1}{2}\left(\frac{100\ \Omega}{1761\ \Omega}\right)^4 \approx 0.9999948\ \ldots$$

This value is just shy of 5σ (≈ 99.99994 %).

For 1 TΩ (main text Fig. 1 (d)):

The $C_5$ symmetric design utilizes a dual-level star-mesh configuration. Each arm is also its own star-mesh configuration which, starting from an external terminal, has 14 Hall elements in series followed both a grounded shunt (10 parallel elements) and an additional 16 Hall elements in series. All five arms share a secondary grounded shunt with 15 elements in parallel. A measurement between any two outer terminals while grounding the central shunt to the same ground as the two arm shunts yields approximately 1 TΩ.

To get $\mu_n$, recall Eq. S5 and use $R_{\text{eq}} = R_{\text{A}} + R_{\text{B}} + \frac{R_{\text{A}}R_{\text{B}}}{R_{\text{main-shunt}}}$, with an arm now containing multiple segments $R_{\text{A}} = R_{\text{seg1}} + R_{\text{seg2}} + \frac{R_{\text{seg1}}R_{\text{seg2}}}{R_{\text{arm-shunt}}}$, and $R_{\text{seg1}} = R_i + 13R_H$ (since, under ideal conditions $R_{\text{seg1}} = 14R_H$, but one of the elements $R_i$ is failing). Assuming the fault is in Arm A in the measurement loop (Arm B will yield an identical analysis):

$$\mu_n = \mu_i \left| \frac{\partial R_{\text{eq}}}{\partial R_{\text{A}}} \times \frac{\partial R_{\text{A}}}{\partial R_{\text{seg1}}} \times \frac{\partial R_{\text{seg1}}}{\partial R_i} \right|$$

$$\mu_n = \mu_i \left[ \left(1 + \frac{2270R_H}{R_H/15}\right) * \left(1 + \frac{16R_H}{R_H/10}\right) * 1 \right] \approx 5.48\ \text{M}\Omega$$

Or, if the fault is in Arm A's second segment (containing 16 elements in series):

$$\mu_n = \mu_i \left[ \left(1 + \frac{2270R_H}{R_H/15}\right) * \left(1 + \frac{14R_H}{R_H/10}\right) * 1 \right] \approx 4.80\ \text{M}\Omega$$

Or, if the fault is in Arm A's shunt (one of the 10 elements in parallel):

$$\mu_n = \mu_i \left[ \left(1 + \frac{2270R_H}{R_H/15}\right) * \left(-\frac{(14R_H)(16R_H)}{(R_H/10)^2}\right) * \left(\frac{1}{10}\right)^2 \right] \approx 7.63\ \text{M}\Omega$$

Or, if the fault is not in an arm, but rather in the central grounded shunt (one of the 15 elements in parallel):

$$\mu_n = \mu_i \left[ \left( -\frac{(2270R_H)(2270R_H)}{(R_H/15)^2} \right) * 1 * \left( \frac{1}{15} \right)^2 \right] \approx 5.15 \text{ M}\Omega$$

We select the smallest value of $\mu_n$ to set a lower bound for the SPQR criterion.

The five-measurement suite can be set up in the following manner (we chose *N* to be 5 to match $n$, the order of rotational symmetry, for keeping a square matrix, but having a larger *N* is now possible and would only improve the SPQR criterion). The sensitivity matrix $\boldsymbol{S}$ reflects the five measurements one may take:

$$\boldsymbol{S} = \begin{bmatrix} 1 & 1 & 0 & 0 & 0 \\ 0 & 1 & 1 & 0 & 0 \\ 0 & 0 & 1 & 1 & 0 \\ 0 & 0 & 0 & 1 & 1 \\ 1 & 0 & 0 & 0 & 1 \end{bmatrix}$$

The determinant evaluates precisely to $\det(\boldsymbol{S}) = 2$. The SPQR criterion is:

$$\Phi \approx 1 - \frac{1}{2}\left(\frac{0.1\ M\Omega}{4.80\ M\Omega}\right)^5 \approx 0.999999998 \ldots$$

This value is stronger than 5σ (≈ 99.99994 %). Taking four measurements on a $C_4$ symmetric design for this value would also be stronger than 5σ.

<u>For $\frac{2}{3}\boldsymbol{R}_{\mathbf{H}}$ QHARS</u> (Fig. 1 (e)):

The $C_3$ symmetric design has three arms where each must have a resistance of about 6.45 kΩ. With 2 Hall elements in parallel for each of the three arms ($L$), one can get the macroscopic breakdown parameter $\mu_n$ by using the fact that $R_{\text{eq}} = R_{\text{arm-A}} + R_{\text{arm-B}}$, $R_{\text{arm-A}} = R_{\text{seg}}$ (since there are no further complications in the topology of each arm), and $R_{\text{seg}} = \left[\frac{L}{R_i}\right]^{-1}$:

$$\mu_n = \mu_i \left| \frac{\partial R_{\text{eq}}}{\partial R_{\text{arm-A}}} \times \frac{\partial R_{\text{arm-A}}}{\partial R_{\text{seg}}} \times \frac{\partial R_{\text{seg}}}{\partial R_i} \right|$$

$$\mu_n = \mu_i \left[ 1 * 1 * (-1) \left(\frac{L}{R_i}\right)^{-2} * (-1) \left(\frac{1}{R_i}\right)^{2} \right]$$

$$\mu_n \approx \frac{\mu_i}{L^2} = \frac{1.00\ \Omega}{2^2} \approx 0.25\ \Omega$$

The sensitivity matrix that contains three row defect states ($\delta_A, \delta_B, \delta_C$) yielding:

$$\boldsymbol{S} = \begin{bmatrix} 1 & 1 & 0 \\ 0 & 1 & 1 \\ 1 & 0 & 1 \end{bmatrix}, \quad |\det(\boldsymbol{S})| = 2$$

To get the observation window, we must use the nominal value of the measurement:

$$\epsilon_{QHARS\,Device} = (12906.4037\ \Omega) \times 5 \times 10^{-8} \Omega/\Omega = 6.45 \times 10^{-4} \Omega$$

Now use the result for the SPQR criterion (Eq. 8 in the main text):

$$\Phi \approx 1 - \frac{1}{2} \left( \frac{6.45 \times 10^{-4} \Omega}{0.25\ \Omega} \right)^3 \approx 0.999999991 \ldots$$

And this result is significantly stronger than 5σ (≈ 99.99994 %).

## 6. Selection of 5σ and the Single-Element Comparison

In the main text, the SPQR criterion analysis proposes a confidence threshold of 5σ, conceptually analogous to the stringent statistical metrics used for major physics discoveries. While this threshold strictly bounds the probability of localized quantization breakdown, combining the SPQR criterion with a standard uncertainty budget, which accounts for external systematic offsets, provides an operational proxy for array quantization confidence in practice.

With the reconciliation of that proposal with the measurements corresponding to the array in the main text, here we explore even further reconciliation with historical practices in measuring single quantum Hall element devices – specifically a measurement in which two orthogonal Hall resistances and two diagonal measurements are performed (as to include a longitudinal resistance along the dissipationless edge of a single Hall element), which harbor their own form of rotational symmetry.

One can make the following definitions and procedure for assessing measurements within the SPQR criterion framework:

First, in a 6-terminal device, current flows from the Source on the left to the Drain on the right. There are four electrical contacts for Hall and diagonal measurements: Top-Left ($V_{TL}$), Top-Right ($V_{TR}$), Bottom-Right ($V_{BR}$), and Bottom-Left ($V_{BL}$). The edge state electron flow on the top and bottom edges moves in opposite directions. Because one is only measuring between these four contacts, the internal defect vector $\vec{\delta}$ consists of just $n = 2$ longitudinal dissipation zones, namely $\delta_T$ and $\delta_B$ (again, one deviation for each edge, where breakdown is set to occur). Therefore:

$$V_{TL} = V_{top}$$

$$V_{TR} = V_{top} - \delta_T$$

$$V_{BR} = V_{bot}$$

$$V_{BL} = V_{bot} - \delta_B$$

Execute the 4 measurements and map their deviations ($\Delta R$) from the ideal quantized Hall resistance ($R_H = V_{top} - V_{bot}$). The 2 orthogonal (Hall) measurements are the left (top two equations) and right (bottom two equations) Hall pairs:

$$V_{TL} - V_{BL} = V_{top} - (V_{bot} - \delta_B) = R_H + \delta_B$$

$$\Delta R_1 = \delta_B$$

$$V_{TR} - V_{BR} = (V_{top} - \delta_T) - V_{bot} = R_H - \delta_T$$

$$\Delta R_2 = -\delta_T$$

And for the first diagonal measurements, we first have $V_{TL}$ to $V_{BR}$:

$$V_{TL} - V_{BR} = V_{top} - V_{bot} = R_H$$

$$\Delta R_3 = 0$$

(Since both probes sit upstream of the defects, this ideally matches $R_H$)

The second diagonal is $V_{TR}$ to $V_{BL}$:

$$V_{TR} - V_{BL} = (V_{top} - \delta_T) - (V_{bot} - \delta_B) = R_H - \delta_T + \delta_B$$

$$\Delta R_4 = -\delta_T + \delta_B$$

The sensitivity matrix and determinant both map our 4 measurements to our 2 defect variables ($\vec{\delta} = [\delta_T, \delta_B]^T$), yielding a $4 \times 2$ rectangular sensitivity matrix $\boldsymbol{S}$:

$$\begin{bmatrix} \Delta R_1 \\ \Delta R_2 \\ \Delta R_3 \\ \Delta R_4 \end{bmatrix} = \begin{bmatrix} 0 & 1 \\ -1 & 0 \\ 0 & 0 \\ -1 & 1 \end{bmatrix} \begin{bmatrix} \delta_T \\ \delta_B \end{bmatrix}$$

We then get the matrix $\boldsymbol{S}^T\boldsymbol{S}$:

$$\boldsymbol{S}^T\boldsymbol{S} = \begin{bmatrix} 0 & -1 & 0 & -1 \\ 1 & 0 & 0 & 1 \end{bmatrix} \begin{bmatrix} 0 & 1 \\ -1 & 0 \\ 0 & 0 \\ -1 & 1 \end{bmatrix} = \begin{bmatrix} 2 & -1 \\ -1 & 2 \end{bmatrix}$$

Evaluating the determinant of this compressed 2 × 2 matrix:

$$\det(\boldsymbol{S}^T\boldsymbol{S}) = (2)(2) - (-1)(-1) = 4 - 1 = 3$$

The SPQR criterion (with $N = 2$) uses $\epsilon_{QHARS\,Device}$ since the output is identical and exemplary for historical metrological practices. For $\mu_n$, we will consider two cases. The first is the proposed quantization breakdown value of 1, yielding:

$$\Phi \approx 1 - \frac{1}{\sqrt{3}}\left(\frac{6.45 \times 10^{-4}\Omega}{1\,\Omega}\right)^2 \approx 0.999999760 \ldots$$

And this result is stronger than 5σ (≈ 99.99994 %), marking a deep compatibility between historical practice and proposed adoption of the SPQR criterion.

## 7. Validation via Longitudinal Resistances of the QHARS Device

Precision measurements of the longitudinal resistances for each of the six elements of the QHARS device were collected using a nanovoltmeter (EM N11, see main text Acknowledgments) and a NIST-built ramping voltage source that is normally used in a CCC resistance bridge system [1]. Current is applied to the full device (either 771 μA or 1157 μA) while the magnetic flux density is ramped along the plateau ($i$ = 2 Landau level) and a potential difference is measured using the auxiliary contact pads in the design for each element (see main text for layout). Though this configuration can be applied for a few-element QHARS device, the engineering of these validating electrical contacts becomes rapidly infeasible as a function of design complexity.

An image of the remounted device is in Fig. 1-SM and the corresponding data of the six individual elements is in Fig. 2-SM and Fig. 3-SM for the two values of applied electrical current.

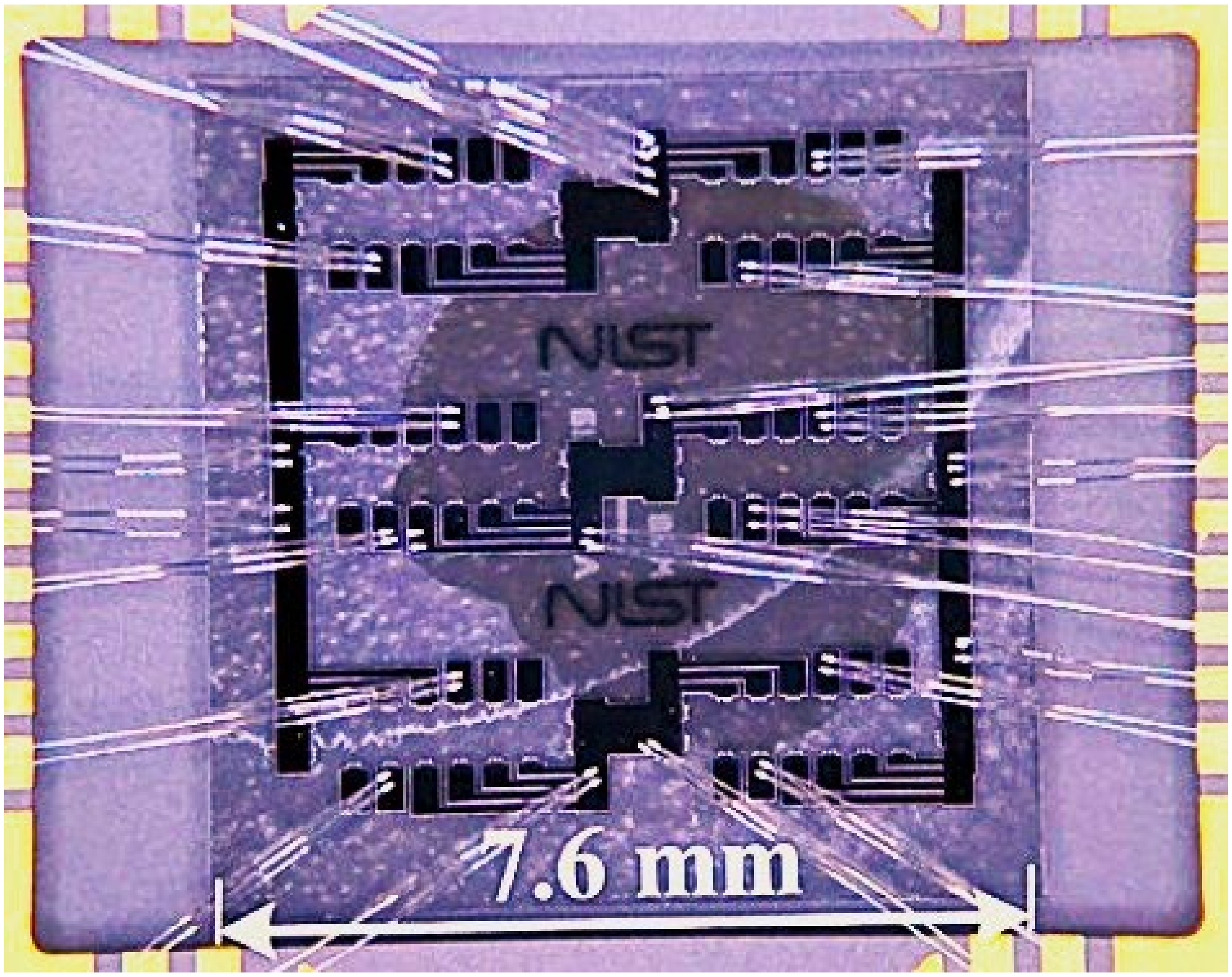

Fig. 1-SM. Optical image of device remounted on a 32-pin chip carrier, specifically to test the longitudinal resistances of each of the six elements.

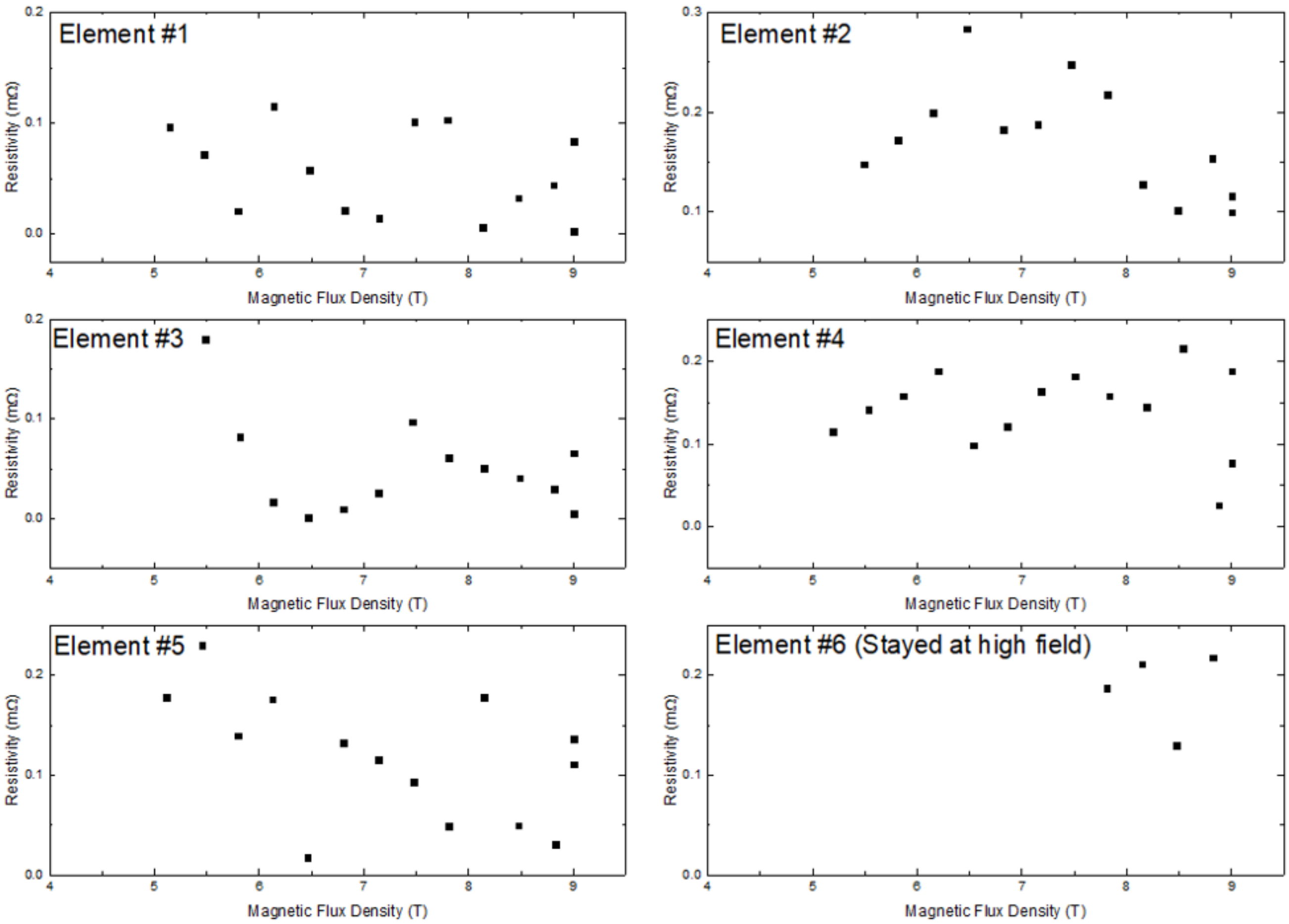

Fig. 2-SM. Longitudinal resistances (converted to resistivities) of each of the six elements. The electrical current was 771 μA.

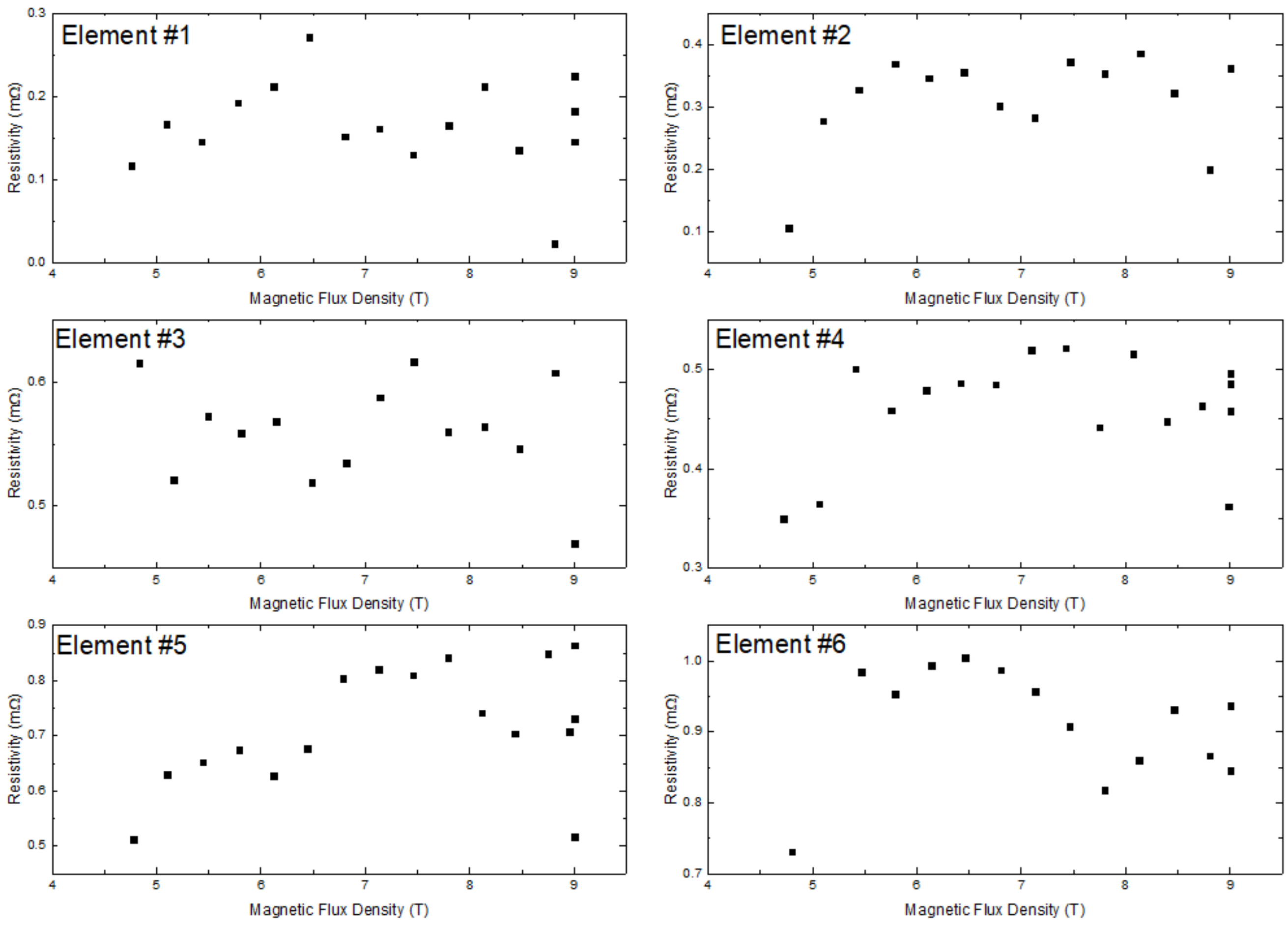

Fig. 3-SM. Longitudinal resistances (converted to resistivities) of each of the six elements. The electrical current was 1.157 mA.

## 8. Possible Device Designs

This section focuses on showing potential design templates for the topologies presented in the main text. Each value has a corresponding design that can be modified to incorporate greater rotational symmetry by introduction of additional elements. Doping control may be obtained be either functionalization (main text) or other methods like *h*-BN coverage for smaller scale graphene areas [2].

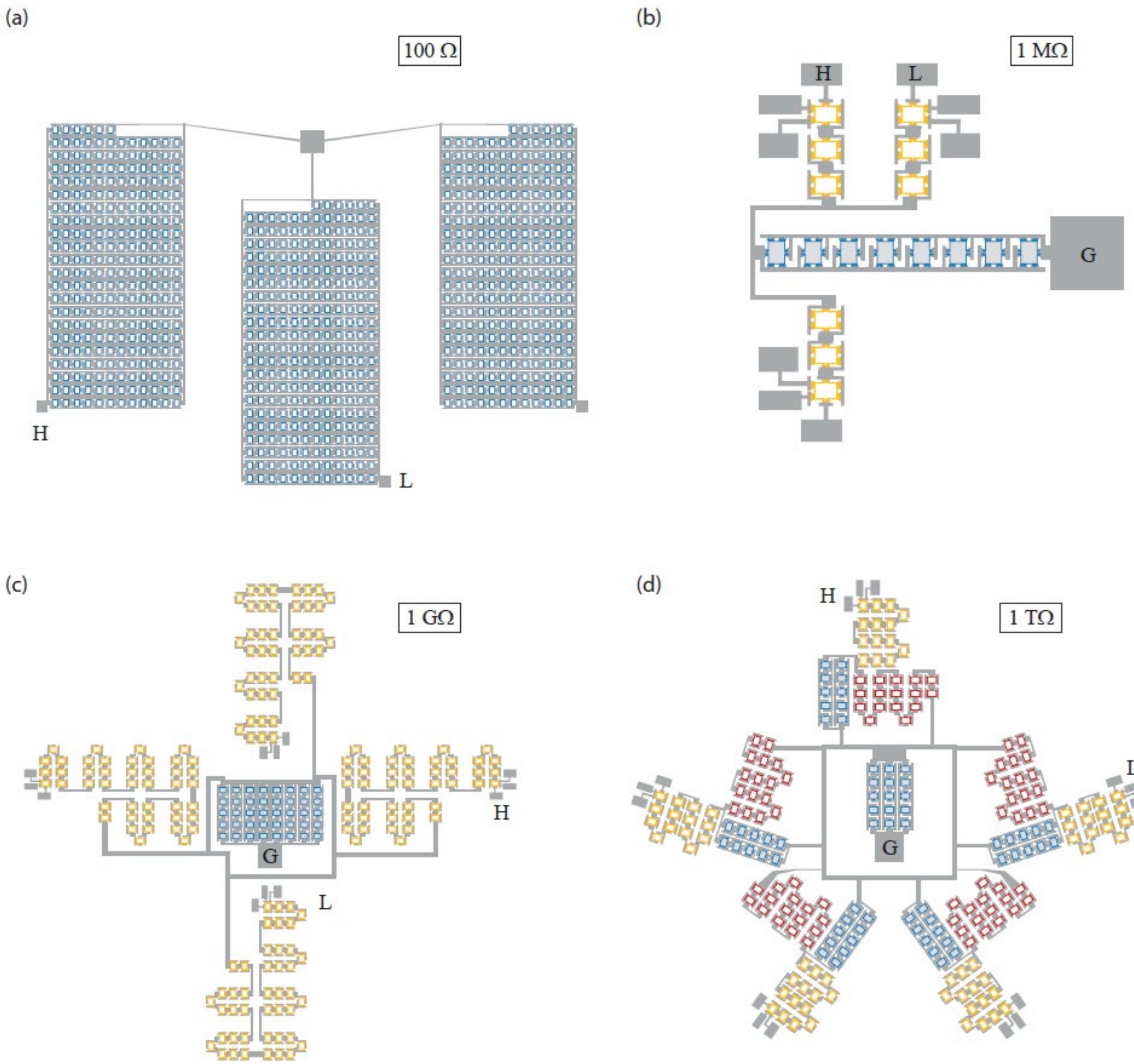

Fig. 4-SM. Possible device designs of the topologies presented in the main text (and color coded to those drawings).